
\input phyzzx
\input psfig
\hoffset=0.2truein
\voffset=0.1truein
\hsize=6truein
\def\TITLEPAGE{\frontpagetrue}
\def\CALT#1{\hbox to\hsize{\tenpoint \baselineskip=12pt
        \hfil\vtop{\hbox{\strut CALT-68-#1}
        \hbox{\strut DOE RESEARCH AND}
        \hbox{\strut DEVELOPMENT REPORT}}}}

\def\CALTECH{
        \address{California Institute of Technology,
Pasadena, CA 91125}}
\def\TITLE#1{\vskip .5in \centerline{\fourteenpoint #1}}
\def\AUTHOR#1{\vskip .2in \centerline{#1}}

\def\ABSTRACT#1{\vskip .2in \vfil \centerline{\twelvepoint
\bf Abstract}
        #1 \vfil}
\def\ENDTITLEPAGE{\vfil\eject\pageno=1}
\def\mapright#1{\smash{\mathop{\longrightarrow}\limits^{#1}}}
\def\hoek{\hbox{\vrule height 2.5ex depth 0pt \vrule width 2.5ex height .4pt
 depth 0pt}}
\def\haak#1#2{
\mathop{\hoek\llap{\vbox to 2.5ex{ \vfil
\hbox{$\scriptstyle#1$\hskip 2.8ex} \vfil}}}
\limits_{#2} }
\def\hook#1#2{\setbox0=\hbox{$\scriptstyle#1$}
\hskip\wd0\haak{\box0}{#2}}
\def\caption#1#2{\vskip 0.1in\centerline{\vbox{\hsize 5in\noindent
     \tenpoint\baselineskip=14pt\strut Fig.~#1: #2\strut}}}
\hfuzz=5pt
\tolerance=10000
\singlespace
\TITLEPAGE
\CALT{1873}
\TITLE{Vortices on Higher Genus Surfaces\foot{Work supported in part
by the U.S. Dept. of Energy under Grant no. DE-FG03-92-ER40701.}}
\AUTHOR{Kai-Ming Lee\foot{\tt email address: kmlee@theory3.caltech.edu}}
\CALTECH

\ABSTRACT{We consider the topological interactions of vortices on general
surfaces. If the genus of the surface is greater than zero, the handles
can carry magnetic flux. The classical state of the vortices and the
handles can be described by a mapping from the fundamental group to
the unbroken gauge group. The allowed configurations must satisfy a
relation induced by the fundamental group. Upon
quantization, the handles can carry ``Cheshire charge.'' The motion of
the vortices can be described by the braid group of the surface.
How the motion of the vortices affects the state is analyzed in detail.}
\ENDTITLEPAGE
\vfill\eject
\normalspace
\chapter{Introduction}

When a gauge symmetry is spontaneously broken, in general
there will be stable topological
\REF\Preskill{J.~Preskill, Vortices and monopoles, {\it in\/} Architecture
     of the fundamental interactions at short distances,
     ed. P.~Ramond and R.~Stora (North-Holland, Amsterdam, 1987).}
defects.\refmark\Preskill\ What types of defects will be created depends on
the spacetime dimension and the topology of the vacuum manifold. In two
spatial dimensions, if the fundamental group of the vacuum manifold is
non-trivial, there will be point defects which are called vortices.
A charged particle winding around a vortex will be transformed by an element
of the unbroken gauge group. This is the
\REF\AB{Y.~Aharonov and D.~Bohm, Phys. Rev. 119 (1959) 485.}
\REF\Schwarz{A.~S.~Schwarz, Nucl. Phys. B208 (1982) 141.}
\REF\KraWil{L.~Krauss and F.~Wilczek, Phys. Rev. Lett. 62 (1989) 1221.}
\REF\Benson{M.~Alford, K.~Benson, S.~Coleman, J.~March-Russell and
     F.~Wilczek, Phys. Rev. Lett. 64 (1990) 1632; Nucl. Phys. B349
     (1991) 414.}
\REF\PreKra{J.~Preskill and L.~Krauss, Nucl. Phys. B341 (1990) 50.}
\REF\Bucher{M.~Bucher, Nucl. Phys. B350 (1991) 163.}
\REF\Disentang{M.~Alford, S.~Coleman and J.~March-Russell, Nucl. Phys.
     B351 (1991) 735.}
\REF\Neworder{M.~Alford and J.~March-Russell, Nucl. Phys. B269 (1992)
     276.}
\REF\qft{M.G.~Alford, K.-M.~Lee, J.~March-Russell and J.~Preskill,
     Nucl. Phys. B384 (1992) 251.}
\REF\BucherLo{M.~Bucher, H.-K.~Lo and J.~Preskill, Nucl. Phys. B386
     (1992) 3.}
\REF\BucherLee{M.~Bucher, K.-M.~Lee and J.~Preskill, Nucl. Phys. B386
     (1992) 27.}
(non-abelian) Aharonov-Bohm effect.\refmark{\AB-\BucherLee}\ It is
long-range and topological. This means that the gauge transformation will not
depend either on how far apart the particle and the vortex
are, or on the exact loops
the charged particle travels along, as long as their linking numbers with
the vortex are the same. We will say that the vortex carries (non-abelian)
magnetic flux.

Another way to look at it is that in the presence of the vortices, the
fundamental group of the surface is non-trivial.\refmark\Bucher\
After a charged particle
travels along a non-contractible loop around a vortex,
it will remain the same only up to a gauge transformation.
The element of the unbroken gauge group associated with that transformation
is the magnetic flux carried by the vortex. However, the fundamental group
of the surface may be non-trivial even without any vortices. For example,
there may be handles on the surface. There are two non-equivalent
non-contractible loops associated to each handle. Then, by the same argument,
we expect we can assign group elements to the two loops and the handle
can carry magnetic flux; therefore, the handles will have topological
interactions with the vortices and the charged particles. If we interchange
two vortices or let a vortex go along a non-contractible loop, the magnetic
flux carried by the vortices and the loop will be changed. This kind
of motion of the vortices can be described by the braid group of the
\REF\Lad{Y.~Ladegaillerie, Bull. Sci. Math. 100 (1976) 255.}
\REF\Imbo{T.D.~Imbo and J.~March-Russell, Phys. Lett. B252 (1990) 84.}
surface.\refmark{\Lad, \Imbo}\
We then have a natural action of the elements of the braid group
on the states of the vortices and the surface.

If the surface is compact, there is one relation between the generators
of the fundamental group of the surface, the group elements assigned
to the vortices and the non-contractible loops must satisfy
a relation induced from that relation.
This restricts the possible magnetic flux carried by the vortices and
handles on any compact surfaces. The simplest
example is that we cannot put a single non-trivial vortex on a sphere.

In the semiclassical approximation, a pair consisting of a vortex and
an anti-vortex may carry electric charge,
``Cheshire charge.''~~\refmark{\Benson, \PreKra, \qft}\
It turns out that the properties of a handle are
similar (but not equal) to the properties of two
vortex-anti-vortex pairs. In particular, a handle can
carry Cheshire charge. If the size of the handle is very small,
an observer outside the handle will see a ``particle'' that carries both
magnetic flux and electric charge, a dyon.\refmark\Preskill\
(The term ``dyon'' is originally for a particle that carries
both electric and magnetic charge in $3+1$ dimensions. We stretch its
meaning to $2+1$ dimensional spacetime.)
In fact, any particle that carries magnetic flux and/or electric charge
\REF\Dijk{R.~Dijkgraaf, V.~Pasquier and P.~Roche, Nucl. Phys. B (Proc.
     Suppl.) 18B (1990) 60.}
\REF\Bais{F.A.~Bais, P.~van~Driel and M.~de~Wild~Propitius,
     Phys. Lett. B280 (1992) 63.}
falls into the representations of the quantum double associated
with the gauge group.\refmark{\Dijk, \Bais}\
In the language of the quantum double, we have a
unified treatment of the magnetic flux and electric charge. There is also
a restriction on the configurations of the dyons on any compact surface.

In section 2, the basic properties of vortices will be briefly reviewed.
The purpose of this section is to establish conventions. In section 3,
the braid group of a surface will be described and the topological
interactions between vortices and handles will be analyzed. We will find
out that, locally, there is no restriction on the assignment of group
elements to the non-trivial loops of a handle.
In section 4, we will give a semi-classical
analysis of the theory. The argument that the handle can carry
Cheshire charge is given. We also explain what a quantum double is
and why it is relevant. In section 5, the most general formulation of
dyons on a surface is described. (The analysis of the previous sections is
a special case in this formulation.) We give the conclusions and some
comments in section 6.

\chapter{Non-abelian vortices}

We assume that for our theory, in the high energy regime, the gauge group
is a simply connected Lie group. In the low energy regime, the symmetry is
spontaneously broken by the Higgs mechanism, say, to a finite group, $G$.
Then in $2+1$ dimensions, the point defects will be
classified by $\pi_0(G) \cong G$.\refmark\Preskill\
{}From now on, we only consider
the unbroken gauge group and its representations. The original
broken gauge group plays no role in the following discussion.
If the energy scale of the symmetry
breaking is very high, the size of the vortices will be very small.
Low energy experiment usually cannot probe the core of vortices.
Then the
space that a low energy particle sees is the original space with
the points where the vortices are removed.

\FIG\vortex{}
We can assign a group element to any isolated vortex to label the flux
by the following method. Choose a fixed but arbitrary base
point, $x_0$ (away from the vortex), and a loop around it.
Then, calculate the untraced Wilson loop,
$$a(C,x_0)=P\exp\left(i\int_{C,x_0}A\cdot dx\right)~,\eqn\flux$$
where $P$ denotes the path ordering. The orientation of the loop,
$C$, is only a convention. We adopt the convention indicated in
Fig.~\vortex. Then, if a charged particle in representation $(\nu)$ of
$G$ is transported along the loop $C$, it will be transformed by
$D^{(\nu)}(a(C,x_0))$:\refmark\PreKra
$$v^{(\nu)} \to D^{(\nu)}(a(C,x_0))v^{(\nu)}~,\eqn\transform$$
where $v^{(\nu)}$ denotes the state of the charged particle.

\midinsert
\centerline{\psfig{file=vortex.eps,height=2in}}
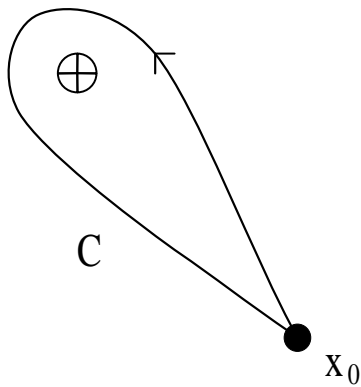
\caption\vortex{The vortex in this figure is represented by a circle with
   a cross. We choose a base point, $x_0$, and a standard path,
   $C$, around the vortex to measure its flux.}
\endinsert

Since the unbroken gauge group is discrete, there is no local low
energy gauge excitation. The group element $a(C,x_0)$ is invariant
under a continuous deformation of the loop $C$. This is how the fundamental
group of the space comes in. We have assigned a group element to a
generator of the fundamental group of the space with punctures.

\FIG\vortices{}
If there are two or more vortices, we have to choose a
standard loop for each vortex as in  Fig.~\vortices.\refmark\Bucher\
Then we can assign group elements to the loops. The combined magnetic
flux, for example, of vortex $1$ and vortex $2$ is the product of
the group elements associated to them. Here we adopt the convention
that the product of two loops, $C_1C_2$, in the fundamental group
means that the particle will travel $C_2$ first and then $C_1$.
So, the combined magnetic flux, in this convention, is $a(C_1C_2,x_0)=
a(C_1,x_0)a(C_2,x_0)$.

\midinsert
\centerline{\psfig{file=vortices.eps,height=2in}}
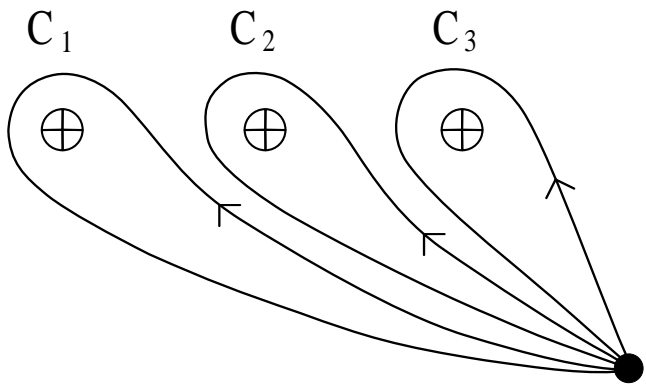
\caption\vortices{For two or more vortices, we choose one standard
   path for each vortex.}
\endinsert

\FIG\intercha{}
Let us consider what will happen if we interchange two vortices.
Let the flux of vortex $1$ and $2$ be $h_1$ and $h_2$ respectively.
If we interchange the vortices counterclockwise,
Fig.~\intercha{}a, the magnetic flux of them will change.
We have to find two loops such that {\it after\/} the interchange,
they will deform to the standard loops. Then the group elements
associated with them are the magnetic flux of the vortices after the
interchange. From Fig.~\intercha{}b, we find that
$$\eqalign{h_1 &\to h_1h_2h_1^{-1} \cr
  h_2 &\to h_1~.\cr}\eqn\interchange$$
We will rely on this kind of loop tracing method to calculate
various processes in the coming sections.

\midinsert
\centerline{\psfig{file=intercha.eps,height=2in}}
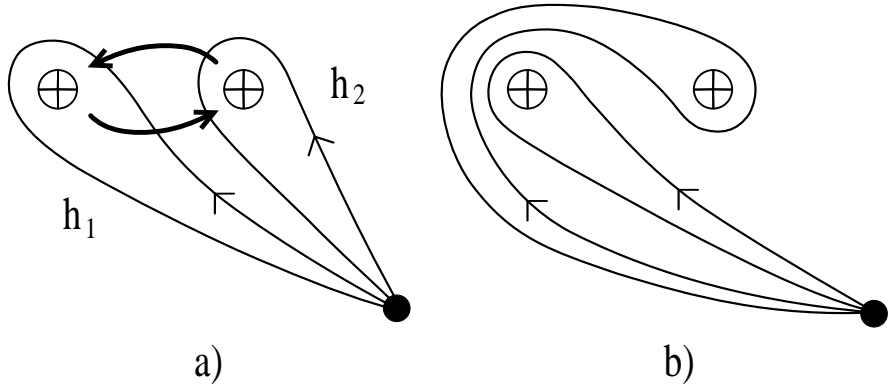
\caption\intercha{In a), the two paths are the standard paths
   based on the same based point. The dark curves with arrows
   represent the interchange of the two vortices. In b), the two
   paths will deform to the standard paths after the interchange
   of the two vortices. So the flux associated with them are
   the flux of the two vortices after the interchange.}
\endinsert

\chapter{Vortices on higher genus surfaces}

\FIG\handle
\FIG\totalflux
\REF\Riemann{See, e.g., H.M.~Farkas and I.~Kra, Riemann Surfaces,
     Springer-Verlag 1980; O.~Forster, Lectures on Riemann Surfaces,
     Springer-Verlag 1981.}
The basic element of an orientable surface with genus greater than zero is
a handle.\refmark\Riemann\
All compact surfaces can be classified according to the
number of handles they have. For a single handle, there are two
generators in the fundamental group. We can choose the generators
to be the loops $\alpha$ and $\beta$ in
Fig.~\handle. Then it is easy to see that the loop in
Fig.~\totalflux\ is equal to $\alpha\beta\alpha^{-1}\beta^{-1}$.

\midinsert
\centerline{\psfig{file=handle.eps,height=2in}}
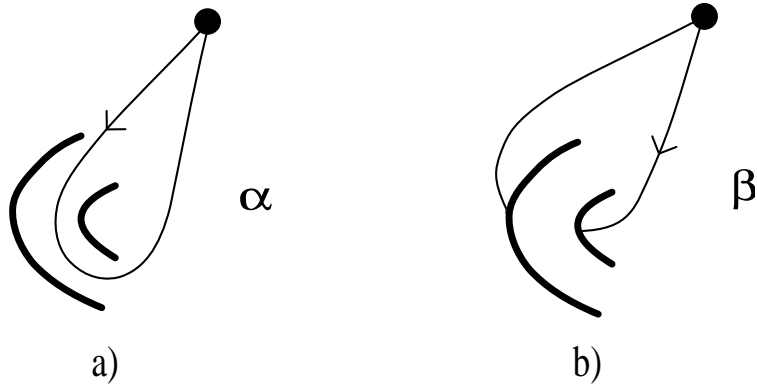
\caption\handle{The wide curves represent a handle that stands out of
   the paper. The two standard paths of the handle are the $\alpha$ in
   a) and $\beta$ in b). Part of $\beta$ is under the handle.}
\endinsert

\midinsert
\centerline{\psfig{file=totalflu.eps,height=2in}}
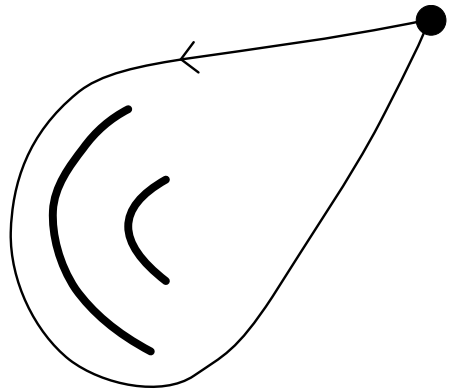
\caption\totalflux{We can use this path to calculate the flux of a
   handle. The path is equal to $\alpha\beta\alpha^{-1}\beta^{-1}$.}
\endinsert

Since $\alpha$ and $\beta$ are non-contractible,
a charged particle transported along them may be transformed by
an element of the gauge group. By sending charged particles
along the loops, we can measure the group elements
associated with them, for example,
$$\eqalign{\alpha &\mapsto a\cr
          \beta &\mapsto b\cr} \eqn\abloop$$
where $a$, $b \in G$.

What will happen if a vortex winds around the loops? We
expect that the magnetic flux of the vortex and group elements
associated with the loops
will change. We can calculate the changes by the loop tracing
method as in section 2. This means that we have to find loops such
that after the traveling of the vortex, these loops deform to
the standard loops we used to measure the flux.

\FIG\alphaloop
\FIG\betaloop
If the vortex with flux $h$ winds around the loop in
Fig.~\alphaloop{}a, we will say that it winds around $\alpha$.
It implicitly means that we have chosen a path (in this case,
the path can be a straight line segment,) from the position of
the vortex to the base point and the vortex goes along this path,
then the $\alpha$ defined in Fig.~\handle\ and finally that
path again in reverse. If also the elements associated with $\alpha$
and $\beta$ are $a$ and $b$ respectively, from Fig.~\alphaloop{}b,
c and d, we find that they will change to
$$\eqalign{h &\to aha^{-1}\cr
           a &\to ahah^{-1}a^{-1}\cr
           b &\to aha^{-1}h^{-1}bah^{-1}a^{-1}~.\cr} \eqn\alphavortex$$
If the vortex winds around $\beta$ in a similar sense, from
Fig.~\betaloop, they will change to
$$\eqalign{h &\to h^{-1}bhb^{-1}h\cr
           a &\to ah\cr
           b &\to h^{-1}bh~.\cr} \eqn\betavortex$$
\midinsert
\centerline{\psfig{file=alphaab.eps,height=2.5in}}
\vskip 0.2in
\centerline{\psfig{file=alphacd.eps,height=2.5in}}
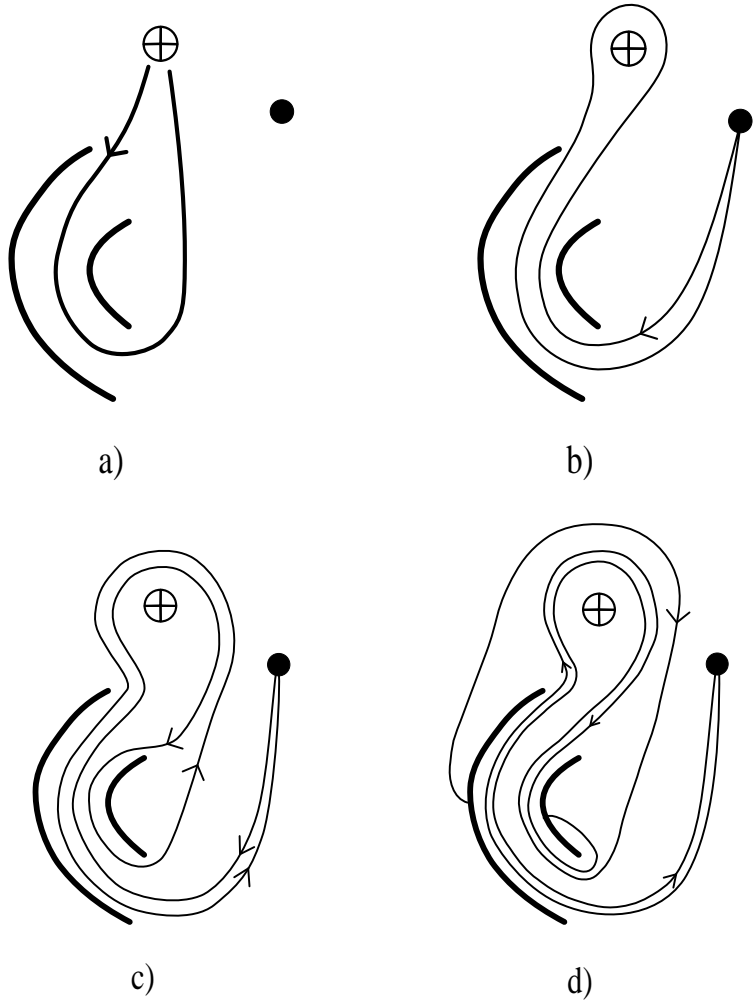
\caption\alphaloop{The motion of the vortex is represented by the loop
   in a). It is also called the $\alpha$ loop. After the motion of
   the vortex as in a), the path in b) will deform to the standard
   path in Fig.~\vortex. The path in c) will deform
   to the standard path $\alpha$ in Fig.~\handle{}a. The path in d) will
   deform to the standard path $\beta$ in Fig.~\handle{}b.}
\endinsert

\midinsert
\centerline{\psfig{file=betaab.eps,height=2.5in}}
\vskip 0.2in
\centerline{\psfig{file=betacd.eps,height=2.5in}}
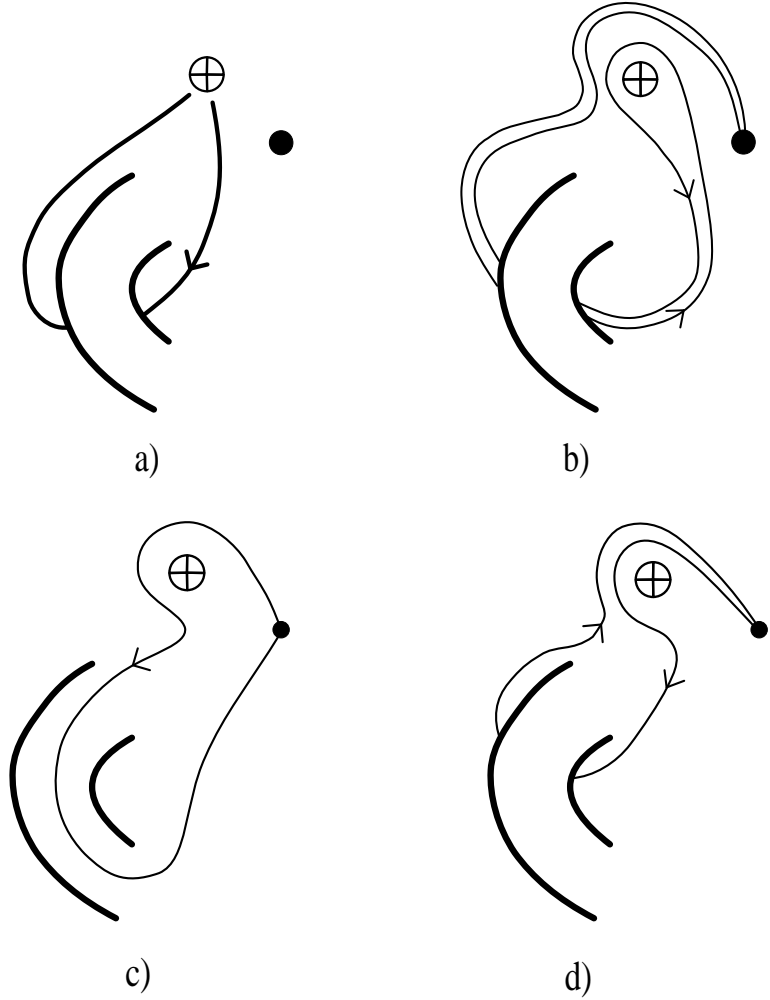
\caption\betaloop{Similar to Fig.~\alphaloop, the loop in a) represents
   the motion of the vortex. Paths in b), c) and d) will deform to
   the standard paths in Fig.~\vortex, $\alpha$ and $\beta$ in
   Fig.~\handle\ respectively.}
\endinsert

Is it possible to assign arbitrary group elements to $\alpha$
and $\beta$? The answer is yes, at least locally. From
\alphavortex\ and \betavortex, it is easy to see that
even if the group is abelian and initially the element associated
with each loop is the identity, after the winding of a non-trivial
vortex, the group elements are no longer trivial. So, we
can transfer the magnetic flux from a vortex to the handle by
sending the vortex to go along $\alpha$ or $\beta$. To excite
the handle to a state with $\alpha\mapsto a$ and $\beta\mapsto b$,
consider the following: We assume that vortices with
arbitrary flux exist and initially the group
elements associated with $\alpha$ and $\beta$ are the identity. And
we send an $ab^{-1}a^{-1}$ vortex to go along $\alpha$. Then, we
send an $a$ vortex to go along $\beta$. By
\alphavortex\ and \betavortex, we have
$$\matrix{\alpha &\mapsto &e &\mapright{\alpha} &e
             &\mapright{\beta} &a\cr
          \beta &\mapsto &e &\mapright{} &aba^{-1}
             &\mapright{} &b&.\cr}\eqn\config$$
Thus, after the vortices execute the prescribed motion, the loops
$\alpha$ and $\beta$ are associated with the desired group elements
$a$ and $b$ respectively.

If the throat of the handle is small or we ignore the
internal structure of the handle and there are no pointlike
vortices hiding inside the handle, we can measure the flux of
the ``particle'' by the loop in Fig.~\totalflux. The flux
is $aba^{-1}b^{-1}$; the flux carried by a handle must
be in this form.

Now let us formulate the theory in precise mathematical terms.
Let the space be an orientable connected surface, $\Sigma$.
If there are $n$ vortices on it, we have to
consider the fundamental group of the
surface with $n$ punctures and a base point,
$\pi_1(\Sigma(n),x_0)$. The combined magnetic flux (or group elements)
of vortices or handles follows from the multiplication rule of
the fundamental group. Any classical state of the vortices and the
surface is a homomorphism from $\pi_1(\Sigma(n),x_0)$ to $G$,
$$\rho : \pi_1(\Sigma(n),x_0) \to G~.\eqn\clastate$$

\FIG\convention{}
If the surface is also compact, there is one relation between
the generators of the fundamental group. For example, the
relation for a surface of genus $g$, $\Sigma_g$, is\refmark\Riemann\
$$\alpha_1\beta_1\alpha_1^{-1}\beta_1^{-1}\ldots
  \alpha_g\beta_g\alpha_g^{-1}\beta_g^{-1} = e~.\eqn\relation$$
For the compact surface of genus $g$ with $n$ punctures,
$\Sigma_g(n)$, the relation is
$$\alpha_1\beta_1\alpha_1^{-1}\beta_1^{-1}\ldots
  \alpha_g\beta_g\alpha_g^{-1}\beta_g^{-1}
  C_1 \ldots C_n = e\eqn\genusrelation$$
where our convention for the loops is shown in Fig.~\convention.

\midinsert
\centerline{\psfig{file=conventi.eps,height=3in}}
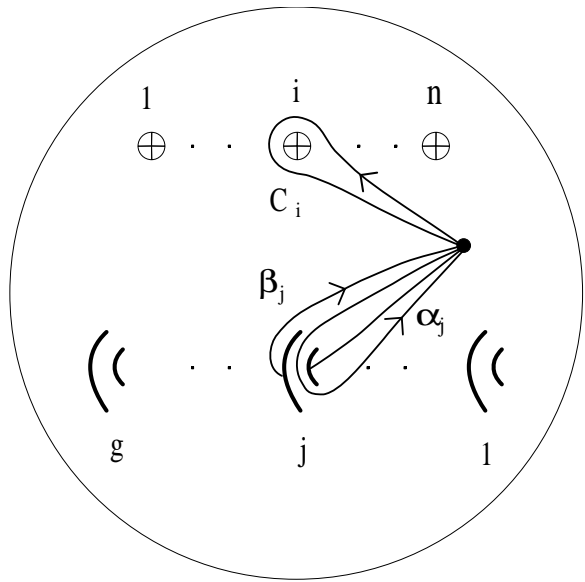
\caption\convention{A compact surface is represented by a large sphere
   with handles here. We choose the standard positions of the $n$ vortices
   and the $g$ handles as in this figure. We also choose a standard
   path for each vortex and the two standard paths $\alpha$ and $\beta$
   for each handle. The positions and the paths are consistent with
   Fig.~\vortices\ and Fig.~\handle.}
\endinsert

The flux of the vortices and handles must satisfy this relation.
For example, the relation associated with a single vortex
on a sphere  is
$$C_1=e~.\eqn\spherelation$$
It is inconsistent to put a single non-trivial vortex on a sphere.
Also, the relation of a torus without any vortex is
$$\alpha_1\beta_1\alpha_1^{-1}\beta_1^{-1}=e~.\eqn\torusrelation$$
Therefore, the group elements associated with the two loops must commute.

This is a global constraint on the possible flux. We have seen
that we can excite the handle to any state locally. We will show
that it is possible to construct the state corresponding to any
homomorphism, $\rho : \pi_1(\Sigma(n),x_0) \to G$. We assume that
we can create vortex-anti-vortex pairs with arbitrary flux; to construct
the state corresponding to $\rho$, we create two sets of vortex-anti-vortex
pairs. The first set consists of $n$ pairs. They contain exactly the $n$
vortices that we want. Then we have $n$ anti-vortices left
and push the anti-vortices to some simply-connected region.
The second set of vortex-anti-vortex pairs contains $2g$ pairs
with appropriate flux.
By sending them to go along the $\alpha$'s and the $\beta$'s, we can excite
the handles to the desired states. After they go along the loops,
their flux will be changed. Now, the combined magnetic flux of the
resulting second set of $4g$ vortices
will no longer be trivial. Let us push them to
the same simply-connected region that contains the $n$ anti-vortices.
We claim that the combined magnetic flux of the first
set of anti-vortices and the second set of vortices is trivial.
Since the surface is compact, a loop wrapped around that region can be
deformed to a loop that wraps around all handles and the
$n$ vortices. The magnetic flux measured along this loop must be
trivial because the flux assigned by $\rho$ satisfies the relation
\genusrelation. This means that the combined magnetic flux of the
left-over vortices is trivial. They can completely annihilate each other
and the state of the surface with the $n$ vortices is given by $\rho$.
We conclude that the space of all states is
${\rm Hom}(\pi_1(\Sigma(n),x_0),G)$.

When we consider the kinematics of the vortices on the surface,
at low energy, they are not allowed to collide with each
other. And because the magnetic flux we measure depends on the
loops we choose, to determine the flux after any motion, the
vortices must be brought back to some standard positions. This kind
of motion is exactly described by the braid group of the surface.

\FIG\braidconven{}
If a collision of particles is not allowed,
the configuration space of $n$ distinguishable particles on a
surface, $\Sigma$, is $\Sigma^n-\Delta$ where $\Delta$ is the
subset of $\Sigma^n$ in which at least two points in the Cartesian
product coincide. The permutation group $S_n$ has an obvious action
on this configuration space. The configuration space of $n$
indistinguishable particles is then $(\Sigma^n-\Delta)/S_n$.
The definition of the braid group of $n$ points on the surface
is\refmark{\Lad, \Imbo}
$$B_n(\Sigma)=\pi_1((\Sigma^n-\Delta)/S_n)~.\eqn\braid$$
If the surface
is the plane, $R^2$, the braid group $B_n(R^2)$ is the usual
braid group with $n-1$ generators which interchange adjacent points.
In $B_n(\Sigma_g)$, there are $2g$ more generators. They are
the $\alpha_i$ and $\beta_i$, $1\le i \le g$ as shown
in Fig.~\braidconven.\foot{We use a different convention from
Ref.~\Lad\ because the $\alpha$'s there involve all handles.}
(We use the same symbols $\alpha$ and $\beta$
to denote the loops in the fundamental group and the generators of
the braid group as explained above \alphavortex.)

These generators are not independent. They satisfy, in our
convention, the following
relations.
$$\eqalign{
\sigma_i\sigma_j&=\sigma_j\sigma_i \qquad |i-j|\ge 2~,\cr
\sigma_i\sigma_{i+1}\sigma_i&=\sigma_{i+1}\sigma_i\sigma_{i+1}
  \qquad 1\le i \le n-2~,\cr
\alpha_1\beta_1\alpha_1^{-1}\beta_1^{-1}\ldots
  &\alpha_g\beta_g\alpha_g^{-1}\beta_g^{-1}
  \sigma_{n-1}\ldots\sigma_1^2\ldots\sigma_{n-1}=e~,\cr
\sigma_i\alpha_l\sigma_i^{-1}\alpha_l^{-1}&=e
  \qquad 2\le i\le n-1{\rm ,}\quad 1\le l\le g~,\cr
\sigma_i\beta_l\sigma_i^{-1}\beta_l^{-1}&=e
  \qquad 2\le i\le n-1{\rm ,}\quad 1\le l\le g~,\cr
\sigma_1\alpha_p\sigma_1^{-1}\alpha_l&=
  \alpha_l\sigma_1\alpha_p\sigma_1^{-1}
  \qquad 1\le p< l\le g~,\cr
\sigma_1\beta_p\sigma_1^{-1}\beta_l&=
  \beta_l\sigma_1\beta_p\sigma_1^{-1}
  \qquad 1\le p< l\le g~,\cr
\sigma_1\alpha_p\sigma_1\alpha_p&=
  \alpha_p\sigma_1\alpha_p\sigma_1
  \qquad 1\le p\le g~,\cr
\sigma_1\beta_p\sigma_1\beta_p&=
  \beta_p\sigma_1\beta_p\sigma_1
  \qquad 1\le p\le g~,\cr
\alpha_p\sigma_1^{-1}\beta_l\sigma_1&=
  \sigma_1^{-1}\beta_l\sigma_1\alpha_p
  \qquad 1\le p<l \le g~,\cr
\beta_p\sigma_1^{-1}\alpha_l\sigma_1&=
  \sigma_1^{-1}\alpha_l\sigma_1\beta_p
  \qquad 1\le p<l \le g~,\cr
\sigma_1\alpha_p\sigma_1\beta_p&=
  \beta_p\sigma_1\alpha_p\sigma_1^{-1}
  \qquad 1\le p\le g~.\cr}\eqn\braida$$

\midinsert
\centerline{\psfig{file=braidcon.eps,height=3in}}
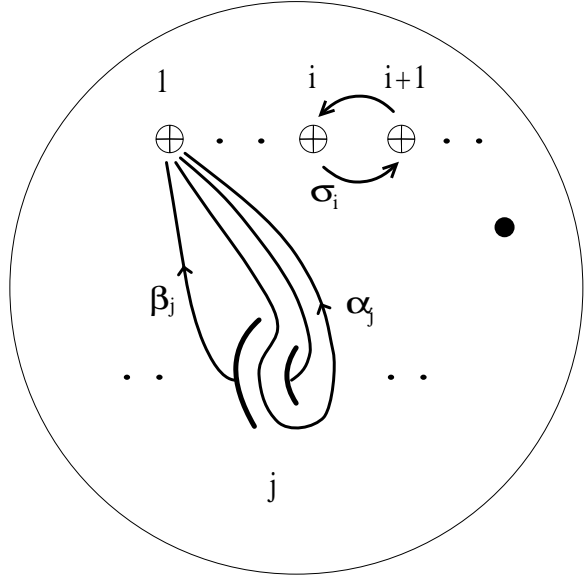
\caption\braidconven{These are the conventions for the braidings of the
   vortices with each other and with the handles. They are equal to
   the braidings in Fig.~\intercha{}a, Fig.~\alphaloop{}a and
   Fig.~\betaloop{}a.}
\endinsert

There is a natural action of the braid group of surface with $n$
points on the fundamental group of the surface with $n$ punctures
defined as follow. If $\tau \in B_n(\Sigma)$ and $\gamma \in
\pi_1(\Sigma(n),x_0)$, define $\tau(\gamma)$ to be the loop such
that {\it after\/} the motion of the $n$ points according to $\tau$,
the loop will deform to $\gamma$. By the calculation in
\interchange, \alphavortex\ and \betavortex, we find that the
nontrivial actions for $B_n(\Sigma_g)$ are
$$\eqalign{\sigma_i(C_i) &=C_{i+1}\cr
           \sigma_i(C_{i+1}) &=C_{i+1}C_iC_{i+1}^{-1}\cr
           \alpha_j(C_1)&=\alpha_jC_1\alpha_j^{-1}\cr
           \alpha_j(\alpha_j)&=\alpha_jC_1\alpha_jC_1^{-1}
	      \alpha_j^{-1}\cr
	   \alpha_j(\beta_j)&=\alpha_jC_1\alpha_j^{-1}
	      C_1^{-1}\beta_j\alpha_jC_1^{-1}\alpha_j^{-1}\cr
           \beta_j(C_1)&=C_1^{-1}\beta_jC_1\beta_j^{-1}C_1\cr
           \beta_j(\alpha_j)&=\alpha_jC_1\cr
	   \beta_j(\beta_j)&=C_1^{-1}\beta_jC_1~.\cr}
	   \eqn\braidaction$$
It is easy to check that this definition satisfies the
relations \braida. This action induces
an action of the braid group on the states of the vortices
and surface. And this is exactly how the state will be changed
after the motion of the vortices.

\chapter{Semi-classical analysis}

We are going to argue that if we specify the flux of $\alpha$ and
$\beta$ of a handle, we know the quantum state of the handle
completely. (Of course, a general quantum state of the handle
could be a linear combination of the flux eigenstates of
$\alpha$ and $\beta$.) The scheme is as follows. We try to find
out a complete set of commuting observables by first choosing
an observable, say $A$, and find out its eigenstates.
In general, there is more than one independent eigenvector
with the same eigenvalue. So, we find another observable, $B$, which
commutes with $A$. Then we can decompose the eigenspaces of $A$
with respect to $B$. If the dimensions of the simultaneous eigenspaces
of $A$ and $B$ are still greater than one, we find yet another
observable which commutes with both $A$ and $B$ and decompose
the eigenspaces and so on. This process will stop if all the
simultaneous eigenspaces are one-dimensional or we run out of
observables.

In our case of discrete gauge theory, there are not
many observables. First of all, the theory is topological. We don't
have any local excitations, and the only things that we can measure
are the magnetic flux and the electric charge bounded by a loop.
For a handle, we can send charged particles to go along $\alpha$
and $\beta$ to measure the flux of them. The measurement of one loop
does not affect the flux of the other, therefore these two observables
commute. Let us denote the state of a handle that $\alpha$ maps to $a$ and
$\beta$ maps to $b$ by $|a,b,X\rangle$ where $X$ specifies any other
quantum numbers needed to completely specify the state of a handle.
Our objective is to prove that no such $X$ is needed.
Now, the only other possible degrees
of freedom, $X$, are the charge bounded by the two loops. It turns out
that we cannot measure the charge bounded by $\alpha$, say, without
messing up the flux of $\beta$. The charge measurement of $\alpha$
does not commute with the measurement of the
flux of $\beta$ and vice versa. Since these
are all the observables in the theory, the flux of $\alpha$ and
$\beta$ form a complete set of commuting observables and we
do not need any $X$.

We now explain why we cannot measure the charge bounded by $\alpha$
without affecting the flux of $\beta$. The only way we can measure
the charge bounded by a loop is to send vortices along the loop and
deduce the charge from the interference pattern.\refmark\Disentang\
(We explain how to measure flux of a vortex by charged particles and
how to measure the charge of a particle by vortices in the Appendix.)
If the handle is in the state $|a,b,X\rangle$ and the flux of the
vortex is $h$, the initial state is $|h\rangle\otimes|a,b,X\rangle$.
Now, suppose that the vortex winds around the loop $\alpha$,
{}from \alphavortex, the final state is $|aha^{-1}\rangle\otimes
|ahah^{-1}a^{-1},aha^{-1}h^{-1}bah^{-1}a^{-1},X'\rangle$ where
the quantum number $X$ may change to $X'$
after the winding of the vortex. The
interference term is $\langle h|aha^{-1}\rangle\langle a,b,X|
ahah^{-1}a^{-1},aha^{-1}h^{-1}bah^{-1}a^{-1},X'\rangle$. The first
factor is non-zero if $a$ and $h$ commute, but then the second
factor is $\langle a,b,X|a,bh^{-1},X'\rangle$ which is zero for
non-trivial $h$; there is no interference and we cannot know
the charge bounded by $\alpha$.

Notice that in some state of the handle, $\alpha$ can bound
some well-defined electric charge. For example, in the state
${1\over\sqrt{|G|}}\sum_{b\in G}|e,b\rangle$, the charge bounded by $\alpha$
can be measured and it is trivial.\foot{I thank Hoi-Kwong Lo
for pointing this out to me.} Some other linear combinations
will give non-trivial charge, however, none of these $\alpha$-charge
eigenstates are $\beta$-flux eigenstates.

\REF\Gcolor{P.~Nelson and A.~Manohar, Phys. Rev. Lett. 50 (1983)
    943; A.~P.~Balachandran, G.~Marmo, N.~Mukunda, J.~S.~Nilsson,
    E.~C.~G.~Sudarshan and F.~Zaccaria, Phys. Rev. Lett. 50 (1983) 1553;
    P.~Nelson and S.~Coleman, Nucl. Phys. B237 (1984) 1;
    A.~P.~Balachandran, F.~Lizzi and V.~G.~Rodgers, Phys. Rev. Lett.
    52 (1984) 1818; P.~A.~Horv\'athy and J.~H.~Rawnsley, J. Math.
    Phys. 27 (1986) 982.}
Recall that if the state of the handle is $|a,b\rangle$,
it carries magnetic flux $aba^{-1}b^{-1}$. In the neighborhood
of such a handle, (in general, in the neighborhood of a vortex with
non-trivial magnetic flux), it is impossible to implement a global
gauge transformation $h$ if $h$ does not commute with the flux of
the handle. That is when we try to extend a local gauge
transformation $h$ along a loop around the handle, the transformation
will be conjugated by the flux of the handle at the end of the loop.
There is no way to solve this inconsistent boundary condition,
called the global color problem.\refmark\Gcolor\ We can only
consider global gauge transformations that are in the normalizer of
$aba^{-1}b^{-1}$, $N_{aba^{-1}b^{-1}}$. Under such a
gauge transformation $h$, the state is transformed to
$$|a,b\rangle \to |hah^{-1}, hbh^{-1}\rangle~.\eqn\statransform$$

Semi-classically, linear combinations of these states are
physically attainable.
The vector space spanned by all these states of a handle can be
decomposed to a direct sum of irreducible representations of
$N_{aba^{-1}b^{-1}}$. These irreducible representations are
the possible Cheshire charges that a handle can carry. Notice
that the mathematical structure of this vector space is
equal to the structure of the states of two
vortex-anti-vortex pairs. They can carry the same kinds of
Cheshire charge.

\REF\Bantay{P.~Bantay, Phys. Lett. B245 (1990) 477; Lett. Math.
     Phys. 22 (1991) 187.}
\REF\Tjin{For a review of quantum doubles and quantum groups,
     see e.g., T.~Tjin, Int. J. Mod. Phys. A7 (1992) 6175.}
Particles that carry both magnetic flux and electric charge are
called dyons. The mathematical tool to classify them is the
quantum double of a group and its
representations. We will
give a brief review of the necessary details here.
Interested readers can look up the references for a full
account.\refmark{\Dijk, \Bais, \Bantay}

The difficulty of classifying dyons is that when a dyon
carries magnetic flux $a$, we can only consider electric charges
which fall into the representations of the normalizer of $a$, $N_a$.
If there are two dyons with flux $a$ and $b$, their electric
charges will be classified by $N_a$ and $N_b$ respectively.
However, when we consider the two dyons as a whole, the total
magnetic flux will be $ab$ (in some convention) and the electric
charge must be a representation of $N_{ab}$. We will find out that
the irreducible representations of the quantum double have exactly
this property. They are labeled by the flux and an irreducible
representation of the normalizer of the flux. A tensor product of two
irreducible representations
can be decomposed to a direct sum of irreducible representations
of the normalizer of the total flux. There is also an element in (the
tensor product of two copies of) the quantum double to implement the
braiding operation.

Let us begin by recalling some properties of representations of
a group. Any representation of a group, $G$, on a vector space,
$V$, is a homomorphism
$$\phi : G \to {\rm End}(V)~.\eqn\grouprep$$
This homomorphism can be extended linearly to the group algebra,
${\cal C}[G]$, by
$$\phi (\sum k_i h_i)= \sum k_i \phi(h_i)~,\eqn\gpalgrep$$
where $k_i\in {\cal C}$. When we consider the tensor product
of two representations, $\phi=\phi_1\times\phi_2$, we have
$$\phi(h)=\phi_1(h)\otimes\phi_2(h)~,\eqn\tensorep$$
if $h$ is a group element. In order to lift to the group
algebra, we define the comultiplication,
$\Delta:{\cal C}[G]\to{\cal C}[G]\otimes{\cal C}[G]$, by
$$\Delta(\sum k_ih_i)=\sum k_ih_i\otimes h_i~.
\eqn\groupcomult$$
Then, $\phi(h)=(\phi_1\otimes\phi_2)\Delta(h)$ where
now $h$ can be any element in the group algebra.

The physical meaning of the comultiplication is that
when a system consists of two subsystems, comultiplication
bridges between the transformation of the whole system and the
individual transformations of the subsystems. In
general, if the symmetry transformations of a theory
form an algebra, we expect there is a corresponding
comultiplication to relate the symmetry transformations
of the whole system and the subsystems.

Now, consider the gauge theory of a finite group, $G$,
in two spatial dimensions. From the above discussion, we
expect to have the following operators. For each
element, $a$, of $G$, there is the gauge transformation
operator of $a$. We can implement this operator by sending an
$a$ vortex around the base point (in some convention). The system
does not change but the basis we used to measure the flux and charge
has changed by a gauge transformation. It is equivalent to relabel
everything in the system. For example, if the flux of a vortex is
initially labeled by $h$, after the transformation, it is labeled
by $aha^{-1}$. We denote this operator by the same
symbol, $a$.

An observer far away from the system can also
measure the total magnetic flux of the system relative to
some fixed gauge choice. We also have a projection
operator, $P_a$, for each $a\in G$, to project to the
subspace of the total flux, $a$. The algebra of
operators, $D(G)$, is generated by $a$ and $P_b$ where
$a$ and $b \in G$.

The multiplication of $a$, $b$ in $D(G)$ is same as the
multiplication in the group. Since $P_a$ is a projection
operator,
$$P_aP_b=\delta_{ab} P_a~.\eqn\projector$$
After a gauge transformation of $a$, the magnetic flux of
the system changes from $b$ to $aba^{-1}$, giving us
$$aP_b=P_{aba^{-1}}a~.\eqn\prodofgaugeproj$$
We have completely determined the algebraic structure of
$D(G)$.

The comultiplication, $\Delta : D(G) \to D(G)\otimes D(G)$,
of elements of $G$ is the one we discussed before
$$\Delta(a)=a\otimes a~.\eqn\hoftcomulta$$
If the system is composed of two subsystems and the magnetic
flux of them is $b$ and $c$, then the flux of the whole
system is $bc$. Conversely, if the total flux is $a$, the
flux of the two subsystems can be any $b$ and $c$ as long
as $bc=a$, giving us
$$\Delta(P_a)=\sum_{{b,c}\atop{bc=a}}P_b\otimes P_c~.
\eqn\hoftcomultb$$
In this equation, we have implicitly assumed some
standard paths are chosen. Then for general elements in
$D(G)$, the comultiplication is\foot{$P_ab$ in Ref.~\Dijk\
and \Bais\ is written as $\hook{a}{b}$.}
$$\eqalign{\Delta(P_ab)&=\Delta(P_a) \Delta(b)\cr
     &=(\sum_{{c,d}\atop{cd=a}}P_c\otimes P_d)
       (b\otimes b)\cr
     &=\sum_{{c,d}\atop{cd=a}}(P_cb)\otimes (P_db)~.\cr}
   \eqn\hoftmultc$$

\FIG\systemsbraid{}

\midinsert
\centerline{\psfig{file=systemsb.eps,height=2.5in}}
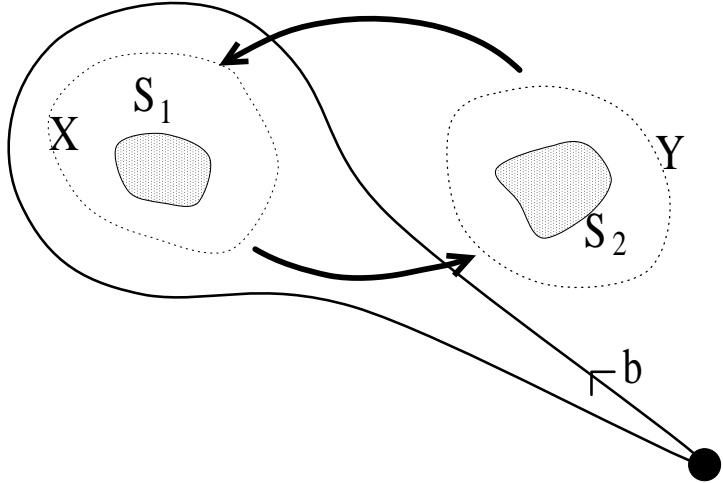
\caption\systemsbraid{This is essentially the same as Fig.~\intercha.
   The shaded areas are the locations of the subsystems. Region $X$ and
   $Y$ are bounded by dotted lines.}
\endinsert

Let us consider the two subsystems, $S_1$ and $S_2$, located
in region $X$ and region $Y$ respectively.
Our convention is that the first factor
in $D(G)\otimes D(G)$ acts on the system in region $X$ and
the second factor acts on system in region $Y$. What will
happen if the two subsystems interchange positions as in
Fig.~\systemsbraid? If the magnetic flux of $S_1$ is $b$,
the effect of the braiding is that apart from the position
change, $S_1$ does not change its state but $S_2$ will be
changed by a gauge transformation $b$,
$$|S_1\rangle\otimes|S_2\rangle\to(b|S_2\rangle)
     \otimes|S_1\rangle~.\eqn\statechangea$$
If $S_1$ is not in a magnetic flux eigenstate, we have
$$|S_1\rangle\otimes|S_2\rangle\to\sum_{b\in G}(b|S_2\rangle)
     \otimes(P_b|S_1\rangle)~.\eqn\statechangeb$$
If we define $\tau$ to be the operator to interchange
the two factors in a tensor product and $R=\sum_bP_b\otimes b
\in D(G)\otimes D(G)$, then the above action can be
described by an operator ${\cal R}=\tau\circ R$ because
$$\eqalign{{\cal R}(|S_1\rangle\otimes|S_2\rangle)
            &=\tau(\sum_{b\in G}(P_b|S_1\rangle)
              \otimes(b|S_2\rangle))\cr
            &=\sum_{b\in G}(b|S_2\rangle)
              \otimes(P_b|S_1\rangle)~.\cr}\eqn\Rop$$
It is easy to show that $R^{-1}=\sum_bP_b\otimes b^{-1}$
and for any $P_ab \in D(G)$,
$$\eqalign{R\Delta(P_ab) R^{-1}&=R(\Delta P_a)(\Delta b)R^{-1}\cr
           &=\sum_{{c,d}\atop {hk=a}}(P_c\otimes c)(P_h\otimes P_k)
                   (b\otimes b)(P_d\otimes d^{-1})\cr
           &=\sum P_cP_hbP_d\otimes cP_kbd^{-1}\cr
           &=\sum P_cP_hP_{bdb^{-1}}b\otimes cP_kbd^{-1}\cr
           &=\sum_{hk=a}P_hb\otimes hP_kb(b^{-1}h^{-1}b)\cr
           &=\sum_{hk=a}P_hb\otimes P_{hkh^{-1}}b\cr
           &=\sum_{hk=a}P_kb\otimes P_hb\cr
           &=\tau(\Delta(P_ab))~.\cr}\eqn\quasitri$$
The meaning of this equation is the following. If an operator
$d\in D(G)$ acts on the whole system, we can calculate its effect on
the subsystems either by directly applying the comultiplication
or by the following procedure. First, interchange the two
subsystems in clockwise direction. Then, apply the
comultiplication and finally, interchange the subsystems
(counterclockwise) back to their original positions.

With the multiplication, comultiplication and the $R$ operator,
(and some other structures) the algebra $D(G)$ is called
the quantum double associated with the group $G$.\refmark\Tjin\ We
have seen that the quantum double is a generalization of
the group algebra and it has direct physical meaning
in a theory with vortices. The particles in such a
theory will fall into representations of the quantum
double.

Now we describe the irreducible representations of
$D(G)$.\refmark\Dijk\
Let the set of all conjugacy classes of $G$ be $\{^AC\}$.
The conjugacy class containing $a$ will be denoted by $[a]$.
For each class, fix an ordering of the elements
$\{^AC\}=\{^Ag_1,\ldots,^Ag_k\}$. Let $^AN$ be the
normalizer of $^Ag_1$. Choose elements, $^Ax_1,\ldots,
^Ax_k \in G$, such that $^Ag_i={}^Ax_i^Ag_1^Ax_i^{-1}$.
We take $^Ax_1=e$. Consider the vector space, $V^A_\nu$,
spanned by the vectors $|^Ag_j,{}^\nu v_i\rangle$,
$j=1$,$\ldots$,k and $i=1$,$\ldots$,dim$\nu$, where $\{{}^\nu v_i\}$
is a basis of the $\nu$ irreducible representation of $^AN$.
This vector space carries an irreducible representation,
$\Pi^A_\nu$, of $D(G)$ defined by
$$\Pi^A_\nu (P_ab)|^Ag_j,{}^\nu v_i\rangle
  =\delta_{a,b^Ag_jb^{-1}}|b^Ag_jb^{-1},
    D^\nu(^Ax^{-1}_lb^Ax_j){}^\nu v_i\rangle\eqn\qdoublerep$$
where $^Ax_l$ is defined by $^Ag_l=b{}^Ag_j b^{-1}$. Notice
that $^Ax^{-1}_lb^Ax_j$ is in $^AN$. The gauge transformation
$b$ is ``twisted'' into the normalizer of the flux.
It can be shown that these representations form a complete set
of irreducible representations of $D(G)$. Any representation
of $D(G)$ can be decomposed into a direct sum of these representations.
In $|^Ag_j,{}^\nu v_i\rangle$, the conjugacy class
labels the magnetic flux of the dyon, the
representation of $^AN$ labels the electric charge.
We can use the comultiplication to define the tensor product
of representations of $D(G)$.

The state of an ordinary electrically charged particle is
$|e,{}^\nu v\rangle$ where now, $\nu$ is an irreducible
representation of $G$. The state of a single
vortex in a group eigenstate is
$|h,1\rangle$ relative to some standard path,
where the $1$ is the trivial representation.
It is found that $\Pi^{[h]}_1\otimes\Pi^{[h^{-1}]}_1
=\Pi^{[e]}_\nu\oplus\cdots$ where $\nu$ is a non-trivial
representation of $G$ and this is the Cheshire charge that
a pair of vortex-anti-vortex can carry.\refmark\Bais\
If we consider the handle in a state $|a;b\rangle$ as a
particle, it has magnetic flux $aba^{-1}b^{-1}$, and the
operator $h$ changes its state to $|hah^{-1},hbh^{-1}\rangle$.
The state of the whole handle has the same transformation
properties under the quantum double as the state
$$|a,1\rangle\otimes|b,1\rangle\otimes
  |a^{-1},1\rangle\otimes|b^{-1},1\rangle~.\eqn\handlestate$$
For example, we can calculate the possible
Cheshire charge of a handle by decomposing the tensor product
$\Pi^{[a]}_1\otimes\Pi^{[b]}_1\otimes\Pi^{[a^{-1}]}_1
\otimes\Pi^{[b^{-1}]}_1$. We must be careful about the meaning
of the expression in \handlestate. It is originally for the
state of four vortices or dyons. In this case, it represents the state of
a single handle. For example, we cannot apply the braiding
operator to it.

\chapter{Dyons on higher genus surfaces}

For any surface, $\Sigma$, with $n$ dyons on it, we
can specify the state by choosing standard paths for
the dyons and the handles and associating a vector in
some representation of the quantum double for each path.
One may expect that there is a correspondence between the
multiplication of paths in the fundamental group and the tensor
product of vectors in representations of the quantum double.
However, the correspondence does not exist. To illustrate this,
consider the product $C_1{C_1}^{-1}$ in Fig.~\vortices. It is
trivial in the fundamental group. The state associated with
it must be the vector in the trivial representation, but if the
state associated with $C_1$ is $|h_1,1\rangle$, the state
associated with $C_1^{-1}$ is $|h_1^{-1},1\rangle$. The tensor
product $|h_1,1\rangle\otimes|h_1^{-1},1\rangle$ transforms as
a linear combination of charge eigenstates, not as the vector
in the trivial representation. The reason why it does not work
is that there is in general no ``inverse'' of a vector in any
representation of the quantum double.\foot{I thank John Preskill
for giving me this example.}

This also occurs in ordinary spacetime. For example, consider
QCD in $3+1$ dimensions. When we say that there are two units of
red charge inside a closed surface, we mean that we have chosen
the {\it outward normal} of the surface and after integrating the
color electric field on the surface relative to this normal
direction, we get two units of red charge. If we consider the
product of the surface and itself with inward normal in the second
homotopy group, and the tensor product of the corresponding charge,
we run into the same difficulty as described above.

However, the tensor product does give us the combined state
of two subsystems. In $3+1$ dimensions, we have to choose the
outward normal (or inward normal) for both surfaces and determine the states
corresponding to these surfaces. Then the state of the combined
system is given by the tensor product. In our case of dyons,
the orientations of the standard loops must be in the ``same sense.''
For example,
if the states associated with $C_1$ and $C_2$ in Fig.~\vortices\
are $|h_1,1\rangle$ and $|h_2,1\rangle$ respectively, the state
associated with $C_1C_2$ is
$|h_1,1\rangle\otimes|h_2,1\rangle$.

For a compact surface, $\Sigma_g$, with $n$ dyons,
we choose the conventions in Fig.~\convention.
The states of the dyons can be measured by charged particles and vortices
traveling around $C_i$ (see Appendix). We can denote the
state where $\alpha_i$ maps to $a_i$, $\beta_i$ maps to $b_i$
and $C_j$ maps to $|h_j,{}^{\nu_j}v\rangle$ by
$$|a_1,b_1,\ldots,a_g,b_g;h_1,{}^{\nu_1}v;
\ldots;h_n,{}^{\nu_n}v\rangle~.\eqn\surfstate$$
A general state will be a linear combination of these.

The state must satisfy the relation discussed in
section 3. This means that for the state
$$\sum_rk_r|a_i^{(r)},b_i^{(r)};h_j^{(r)},
{}^{\nu_j}v_{(r)}\rangle~,\eqn\gsurfstate$$
where the $k_r$'s are constants, the tensor product
$$\eqalign{\sum_r&k_r|a_1^{(r)},1\rangle\otimes|b_1^{(r)},1\rangle
  \otimes|(a_1^{(r)})^{-1},1\rangle\otimes|(b_1^{(r)})^{-1},1\rangle\cr
  &\quad\otimes\ldots\otimes|a_g^{(r)},1\rangle
  \otimes|b_g^{(r)},1\rangle\otimes|(a_g^{(r)})^{-1},1\rangle
  \otimes|(b_g^{(r)})^{-1},1\rangle\cr
  &\quad\otimes|h_1^{(r)},{}^{\nu_1}v_{(r)}\rangle\otimes
  \ldots\otimes|h_n^{(r)},{}^{\nu_n}v_{(r)}\rangle\cr}
  \eqn\gsurfrelation$$
must transform as the trivial representation.
(cf. \genusrelation)

We can also consider the motion of the dyons. Similar
to the discussion in section 3, there is an action of
the braid group on the states of the surface with dyons.
For $B_n(\Sigma_g)$, the action of the $\sigma$'s are
given by the $\cal R$ operator as discussed above.\refmark\Bais\
{}From \braidaction, we also have
$$\eqalign{\alpha_j|a_j,b_j;h_1,{}^{\nu_1}v\rangle
       &=|a_jh_1a_jh_1^{-1}a_j^{-1},
          a_jh_1a_j^{-1}h_1^{-1}b_ja_jh_1^{-1}a_j^{-1}\rangle\cr
        &\qquad\otimes\Pi^{[h_1]}_{\nu_1}(a_j)|h_1,{}^{\nu_1}v
          \rangle\cr
           \beta_j|a_j,b_j;h_1, v^{(\nu_1)}\rangle
             &=|a_jh_1,h_1^{-1}b_jh_1\rangle\otimes
                \Pi^{[h_1]}_{\nu_1}(h_1^{-1}b_j)
                 |h_1,{}^{\nu_1}v\rangle~.\cr}
	   \eqn\qbraidaction$$
One can also check that this
definition satisfies the relations \braida.
Notice that the action of the $\alpha$'s and $\beta$'s cannot
be written as the action of some element of the quantum
double because the flux of $\alpha$ and $\beta$
are not just conjugated. This represents the fact that if
we do see the internal structure of the handle, its state
is not in the Hilbert space of the states of a particle.
If every dyon does not carry electric charge, the representations
$\nu_i$ are trivial. All formulae here then reduce to the
corresponding formulae in section 3.

Let us consider an example. Suppose the group $G$ is the
quaternion group $Q=\{\pm1,\pm i,\pm j,\pm k\}$. There are
five conjugacy classes: $\{1\}$, $\{-1\}$, $\{\pm i\}$,
$\{\pm j\}$ and $\{\pm k\}$. And there are four one-dimensional
irreducible representations: the trivial representation $1$ and
$1_x,1_y,1_z$ where in $1_x$, say, $\pm 1, \pm i$ are represented
by $1$ and the others are represented by $-1$. There is also a
two-dimensional irreducible representation. Notice that the
normalizer of $-1$ is the whole group. So, if a dyon has
flux $-1$, its electric charge can be labeled by representation
of the whole group.

Assume the space is a torus and there are two vortices and one
charged particle. A possible state is
$$\eqalign{|v\rangle={1\over 2}&(|i,j;k,1;k,1;1,1_x\rangle
    +|i,-j;-k,1;-k,1;1,1_x\rangle\cr
    &\quad -|-i,j;-k,1;-k,1;1,1_x\rangle
    -|-i,-j;k,1;k,1;1,1_x\rangle)~.\cr}\eqn\torustate$$
In each term, the first two factors label the flux
carried  by $\alpha$ and $\beta$ ($\pm i$ and $\pm j$).
The third and fourth factors
are the flux ($\pm k$) and the charge (trivial) of the first vortex.
The next two factors have the same meaning. The final
two are the trivial flux ($1$) and the charge ($1_x$) of the
charged particle. The state of the handle and the states of
the vortices are entangled but if we consider them as
a whole, they are in the state $|1,1_x\rangle$,
so together with the charged particle, they satisfy
\gsurfrelation.

If the first vortex winds around $\beta$, by
\qbraidaction, we have
$$\eqalign{\beta|v\rangle=
    &{1\over2}(|-j,-j;k,1;k,1;1,1_x\rangle
    +|j,j;-k,1;-k,1;1,1_x\rangle\cr
    &\quad -|-j,-j;-k,1;-k,1;1,1_x\rangle
    -|j,j;k,1;k,1;1,1_x\rangle)\cr
   =&{1\over2}(|-j,-j\rangle-|j,j\rangle)\otimes
     (|k,1\rangle\otimes|k,1\rangle-|-k,1\rangle\otimes|-k,1\rangle)
      \otimes|1,1_x\rangle~.\cr}\eqn\betatorus$$
Now, the first factor in the above tensor product is the state of
the handle, the second factor is the state of the two vortices and
the last factor is the state of the charged particle. The handle
carries magnetic flux $-1$ and Cheshire
charge $1_y$. The two vortices together carry flux $-1$ and
charge $1_z$. There is magnetic flux transfer between the handle and
the pair of vortices.

What will happen if a charged particle winds around a loop of
a handle? As we have seen in the beginning of section 4, if
the handle is in flux eigenstate, the state of the particle
will be transformed by the flux of the loop and the state of
the handle will remain the same. If the handle is in some
linear combination of flux eigenstates, something interesting
will happen. For example, let the state of the charged particle
be $|v\rangle$, and assume initially the state of the handle is
${1\over\sqrt{|G|}}\sum_{b\in G}|e,b\rangle$. Then the charge bounded by
$\alpha$ is trivial. If the charged particle winds
around $\beta$, then
$$|v\rangle\otimes{1\over\sqrt{|G|}}\sum_{b\in G}|e,b\rangle\to
  {1\over\sqrt{|G|}}\sum_{b\in G}|D(b)v\rangle
  \otimes|e,b\rangle~.\eqn\chargewormhole$$
If we now introduce a $h$ vortex, and wind it around
$\alpha$, the state changes to ${1\over\sqrt{|G|}}\sum_{b\in G}|D(b)v\rangle
\otimes|e,bh^{-1}\rangle$, then the interference term is
proportional to
$$\eqalign{&\sum_{b,b'\in G}\langle e,b'|
  \langle D(b')v|D(b)v\rangle|e,bh^{-1}\rangle\cr
  =&\sum_{b,b'\in G}\langle v|D({b'}^{-1}b)|v\rangle
    \delta_{b',bh^{-1}}\cr
  =&\sum_{b\in G}\langle v|D(h)|v\rangle~.\cr}\eqn\chargemouth$$
\REF\LoLee{H.-K.~Lo and K.-M.~Lee, Wormholes and Charge Conservation,
Caltech preprint CALT-68-1880 (1993).}
The charge bounded by $\alpha$ is $v$ and the flux of $\alpha$
is identity. We see that the charge of the particle is transferred
to the ingoing mouth of the handle. However, the state of the
particle entangles with the state of the handle and can no longer be
specified by a single vector. This kind of charge
transfer between mouths of wormholes or handles and charged particles
also occurs in $3+1$ dimensions and for continuous gauge
groups.\refmark\LoLee

\chapter{Conclusion}

\REF\Brekke{L.~Brekke, H.~Dykstra, S.J.~Hughes and T.D.~Imbo,
     Phys. Lett. B288 (1992) 273.}
\REF\BaisChern{F.~A.~Bais, P.~van~Driel and M.~de~Wild~Propitius,
     Nucl. Phys. B393 (1993) 547.}
We have argued that in $2+1$ dimensions, non-trivial
topology, the handle, can carry magnetic flux classically.
If the unbroken gauge group is finite, we can actually
assign arbitrary group elements to the two non-equivalent
loops associated to the handle.
Semi-classically, the state of a handle can be specified by
the flux of the two non-trivial loops. It can also carry Cheshire
charge. On the other hand, a general particle will fall into
representations of the quantum double, an algebra constructed from
the gauge group. We have also explained the physical meanings of
the elements of the quantum double, the comultiplication and the
braiding operator $\cal R$. If the surface is compact, there
is a relation between the generators of the fundamental group of
the space. This relation restricts the possible flux and charges
of the handles and the dyons on that surface. If the surface is
non-compact, no such relation exists.

There is topological interaction between the dyons and
the handles. The motion of the dyons is described by the
braid group of the surface. Then, the topological interaction
can be described by an action of the braid group on the states
of the handles and the dyons. This action, and hence the
topological interaction, can be completely determined
by the path tracing method explained in section 2.
A similar classical analysis in $3+1$
dimensions for cosmic strings has been done by
Brekke et. al.,\refmark\Brekke and
the classification of dyons has been generalized to theories with
Chern-Simons terms.\refmark\BaisChern

\appendix

In this appendix, we will recall how to measure the flux of a beam of
identical vortices using charged particles\refmark\Disentang\ and how to
measure
the charge of a beam of identical charged particles using vortices.

Assume that we have a beam of identical vortices with unknown flux, $h$,
and we have charged particles in any desired states. We can send the
charged particles in a particular state around the vortices and then
observe the interference patterns. If the state of the particles is
$|v\rangle$ in some representation $\nu$, the interference gives us
$$\langle v|D^{(\nu)}(h)|v\rangle~.\eqn\interference$$
If we replace $|v\rangle$ by $|v\rangle +\lambda|w\rangle$ in the above
equation where $\lambda$ is an arbitrary complex number and subtract
$\langle v|D^{(\nu)}(h)|v\rangle +\langle w|D^{(\nu)}(h)|w\rangle$
{}from it, we have
$$\lambda\langle v|D^{(\nu)}(h)|w\rangle + \lambda^*\langle w|
  D^{(\nu)}(h) |v\rangle~.\eqn\interferencetwo$$
Put $\lambda$ to be $i$ and $1$ in succession, we know the values
of two expressions. A linear combination of them gives us
$$\langle v|D^{(\nu)}(h)|w\rangle~.\eqn\interferencethree$$
We can now determine $\langle v|D^{(\nu)}(h)|w\rangle$ for
arbitrary $|v\rangle$, $|w\rangle$
and $\nu$ and hence the matrix representation
of $h$. If we choose $\nu$ to be some faithful representation,
we can determine $h$.

Now assume that we have a beam of charged particles in some
unknown state, $|v\rangle$, in some unknown irreducible representation,
$\nu$, and we have vortices with any desired flux.
We also assume that $\langle v|v\rangle = 1$.
Then a similar interference experiment will give us
$$\langle v|D^{(\nu)}(h)|v\rangle\eqn\interferencefour$$
for arbitrary $h$. Because $\nu$ is irreducible, the vectors,
$D^{(\nu)}(h)|v\rangle$ for $h\in G$, will span the whole
representation space. We know the inner products of these vectors
because $\langle D^{(\nu)}(h_1)v|D^{(\nu)}(h_2)v\rangle =
\langle v|D^{(\nu)}(h_1{}^{-1}h_2)|v\rangle$. By Gram-Schmidt's
orthogonalization, we can form a basis,
$\{|e_i\rangle :i=1,\ldots d\}$, such that $|e_1\rangle=|v\rangle$
and
$$|e_i\rangle=\sum_{h\in G}c^i_h|D^{(\nu)}(h)v\rangle~.\eqn\linearbasis$$
Notice that the coefficients, $c^i_h$, depend only on the numbers
$\langle v|D^{(\nu)}(h')|v\rangle$. We also have $|D^{(\nu)}(h)v\rangle
=\sum_i b^h_i |e_i\rangle$ for some coefficients $b^h_i$.

Now we have a basis, so we can calculate the character of the
representation and hence determine the representation itself.

Suppose that there is another vector, $|w\rangle$, in the
same representation space such that for all $h$ in $G$,
$\langle w|D^{(\nu)}(h)|w\rangle=\langle v|D^{(\nu)}(h)|v\rangle$.
We are going to prove that $|w\rangle$ is equal to $|v\rangle$
up to a phase. If this can be done, we can uniquely determine the state
of the beam of charged particles by only sending vortices around them and
observing the interference pattern.

Let $|e'_i\rangle=\sum_{h\in G}c^i_h|D^{(\nu)}(h)w\rangle$.
Since the coefficients, $c^i_h$, depend only on the numbers
$\langle v|D^{(\nu)}(h')|v\rangle=\langle w|D^{(\nu)}(h')|w\rangle$,
$|e'_i\rangle$ form a basis. Then there is an operator $L$ such that
$|e'_i\rangle=L|e_i\rangle$. We claim that
$LD^{(\nu)}(h)=D^{(\nu)}(h)L$ for all $h$ in $G$.
First of all, we have $|w\rangle=|e'_1\rangle=L|e_1\rangle=|v\rangle$.
Then, $D^{(\nu)}(h)L|v\rangle=D^{(\nu)}(h)|w\rangle=
\sum_ib^h_i|e'_i\rangle=\sum_ib^h_iL|e_i\rangle=
L\sum_ib^h_i|e_i\rangle=LD^{(\nu)}(h)|v\rangle$.
We have
$$\eqalign{D^{(\nu)}(h)L|e_i\rangle&=D^{(\nu)}(h)|e'_i\rangle
=\sum_{h'\in G}c^i_{h'}D^{(\nu)}(hh')|w\rangle\cr
&=\sum_{h'\in G}c^i_{h^{-1}h'}D^{(\nu)}(h')|w\rangle
=\sum_{h'\in G}c^i_{h^{-1}h'}D^{(\nu)}(h')L|v\rangle\cr
&=L\sum_{h'\in G}c^i_{h^{-1}h'}D^{(\nu)}(h')|v\rangle
=LD^{(\nu)}(h)|e_i\rangle~.\cr}\eqn\preSchur$$
This proves the claim. Since $\nu$ is irreducible,
by Schur's lemma, $L$ is the product of a constant and the
identity operator and $|w\rangle= e^{i\theta}|v\rangle$ because
$\langle w|w\rangle=\langle v|v\rangle=1$.

For a beam of dyons, we can first measure the magnetic flux by
sending charged particles around them. After we know the flux,
we can measure their charge by using vortices with flux which
commutes with the flux of the dyons. Then, we completely
determine the state of the dyons in some representation of the
quantum double.

This analysis can be generalized to the measurement of a single
particle or particles in a reducible representation (with some
limitation).\refmark\Disentang\

\vskip\chapterskip
\titlestyle{Acknowledgments}
I would like to thank Hoi-Kwong Lo and especially John Preskill
for very useful discussions.

\refout
\bye